\documentclass[12pt,preprint]{aastex}

\newcommand{\tbsp}{\rule{0pt}{18pt}}

\slugcomment{Submitted to ApJ}

\shorttitle{A multiwavelength study of the nucleus of NGC821}
\shortauthors{Pellegrini et al.}

\begin{document}

\title{A deep $Chandra$, VLA and $Spitzer$ IRAC study\\
of the very low luminosity nucleus
of the elliptical NGC~821}

\author{S. Pellegrini$^1$, A. Siemiginowska$^2$, G. Fabbiano$^2$,
 M. Elvis$^2$, L. Greenhill$^2$, R. Soria$^{2,3}$, A. Baldi$^2$,
D.W. Kim$^2$ }

\affil{$^1$ Astronomy Department, Bologna University, Italy;
silvia.pellegrini@unibo.it \\ 
$^2$ Harvard-Smithsonian Center for Astrophysics, 60 Garden St, 
Cambridge, MA 02138 \\
$^3$ Mullard Space Science Laboratory, University College London, Holmbury St.
Mary, UK}

\begin{abstract}

The relatively nearby (distance=24.1 Mpc) elliptical galaxy NGC821
hosts an extreme example of a quiescent central massive black hole,
for which deep $Chandra$ observations revealed a nuclear source for
the first time, with $L_{2-10\,{\rm keV}}/L_{Edd}\sim 
10^{-8}$.  We present here a multiwavelength study of this nucleus,
including VLA observations that detect a radio counterpart to the
$Chandra$ nuclear source at 1.4 GHz, with a flux density of 127 $\mu$Jy
and possibly a flat spectral shape; we also consider new $Spitzer$
IRAC observations and archival $HST$ images. With these data we
discuss possible scenarios for the accretion modalities of the sole
material that seems available for fuelling, i.e., the stellar mass
losses steadily replenishing the circumnuclear region.  The final
stages of accretion could be radiatively inefficient and coupled to a
compact nuclear jet/outfow.  The stellar mass losses could instead end
up in a standard disc only if a Compton-thick AGN is
present.  Two extended sources detected by $Chandra$ close to the
nucleus could be due to several unresolved knots in a jet.  If a jet
is present, though, its kinetic energy would be only a very small
fraction of the energy associated with the rest mass of the material
being accreted.  Starformation close to the nucleus is not shown
by the available data. Deeper NICMOS, radio and far-IR observations
are expected to further constrain the accretion process in this
intriguing nucleus.

\end{abstract}


\section{Introduction}\label{intro}

The connection between the mass of a central supermassive black hole
(hereafter MBH) and the properties of the host galaxy (see, e.g., the
$M_{BH}-\sigma$ relation; Ferrarese \& Merritt 2000, Gebhardt et
al. 2000), together with the recognition that a MBH resides at the
center of every massive spheroid in the local universe (e.g.,
Ferrarese \& Ford 2005), has led to the idea that the  
formation and evolution of the MBH and the host
galaxy are intimately related (e.g., Springel,
Di Matteo \& Hernquist 2005, Sazonov et al.  2005, Hopkins et
al. 2006a). In this picture, feedback from the MBH has been advocated
to regulate starformation at early epochs, producing the observed
galaxy mass function and colors (e.g., Croton et al. 2006); in later
epochs, feedback is required to heat the galactic flows solving the
"cooling flow" problem (e.g., Omma et al. 2004; Ostriker \& Ciotti
2005, 2007; Churazov et al. 2005, McNamara et al. 2005).

Although feedback is a promising mechanism for solving many
observational puzzles, our understanding of how accretion and feedback
work, and of their evolution over cosmic time, is still uncomplete. It
has been suggested that after the bright AGN phase accretion switches
to a more radiatively quiet or inefficient mode (e.g., Churazov et
al. 2005, Hopkins et al. 2006b) possibly associated with radio jets or
outflows that may carry out most of the accretion power (Di Matteo et
al. 2003, Pellegrini et al. 2003, Falcke et al. 2004, Allen et
al. 2006). Indeed, the nuclei of most of the nearby spheroids are
either radiatively quiescent or exhibit low levels of activity (e.g.,
Ho 2005, Pellegrini 2005a). Typically, these nuclei are not associated
with detectable radio sources, but the absence of radio emission in
itself does not exclude nuclear activity, since 'frustrated' or not
collimated nuclear outflows would be hard to detect (Nagar et
al. 2005). Also, intermittent AGN activity is suggested by the X-ray
and radio morphology of nearby clusters of galaxies (e.g., Fabian et
al. 2006) and galaxies (e.g., Machacek et al. 2006), and may originate
in an accretion flow that switches between different accretion modes
(Janiuk et al.  2004).

In nearby spheroids we can attempt to constrain observationally the
modalities of accretion (and feedback) through detailed studies of the
stellar population, the ISM and the current star formation rate for
the central $\sim $one hundred parsec region, especially if the AGN emission
does not dominate over the other nuclear components. NGC~821, an
isolated elliptical galaxy at a distance of 24.1~Mpc
(Table~\ref{mainlog}), is an ideal target for studying how these
phenomena have evolved in the local universe. There is compelling
evidence of the presence of a central MBH in this galaxy, from
resolved dynamical studies (Gebhardt et al.  2003).  NGC821 has a very
regular and smooth optical appearance (Lauer et al. 2005) and the old
and metal rich stellar population typical of elliptical galaxies
(Proctor et al. 2005); no cold (HI) or dusty ISM (Sarzi et al. 2006,
Ravindranath et al. 2001) have been observed in it, restricting the
potential reservoir for accretion.

The MBH of NGC821 has a mass of $8.5 \times 10^7 M_{\odot} $
(Table~\ref{mainlog}) and an Eddington luminosity of $L_{Edd}\sim
1.1\times 10^{46}\, \rm erg~s^{-1}$. Therefore, if radiatively
efficient accretion were taking place, this galaxy should be a
luminous AGN.  Instead, this MBH is extremely quiescent. Very low
limits on optical emission lines (H$\alpha$, H$\beta$, or [OIII]) are
reported for the nucleus of NGC821 (Ho 2002; Ho et al. 2003; Sarzi et
al. 2006) and the limits on nuclear radio emission are also
stringent: $< $1.5 mJy at 15 GHz (with 150 mas resolution; Nagar et
al. 2005) and $< $0.5 mJy at 5 GHz (with 5$^{\prime\prime}$
resolution; Wrobel \& Heeschen 1991).

Thanks to a deep $Chandra$ pointing for a total exposure of 230 ksec
(discussed in Pellegrini et al. 2007, hereafter P07), a source was
detected for the first time at the nucleus of NGC821 (called S2).  S2
is extended (Tab.~\ref{tabspec}), but it has a flat
($\Gamma=1.49^{+0.14}_{-0.13}$) unabsorbed X-ray spectrum, excluding a
central concentration of hot, optically thin ISM. Its 2--10 keV luminosity
is $3.8\times 10^{38}$ erg s$^{-1}$. An upper limit
of 2.8$\times 10^{38}$ erg s$^{-1}$ was placed on a possible pointlike
0.3--8 keV emission associated with the MBH; this is one of the smallest
values obtained with $Chandra$ for galactic nuclei (Pellegrini 2005b).

In addition to the nucleus, also the hot ISM that could provide a source of
fuel for the MBH can be measured in the X-rays (e.g., Loewenstein et
al. 2001; Soria et al. 2006a). Our deep, sub-arcsecond $Chandra$
pointing revealed diffuse emission in the central part of NGC~821, but
also led to the detection of a population of X-ray binaries which can
account for most -- if not all -- of the diffuse emission, so that we
could place a very stringent upper limit on the contribution of a hot
gaseous component (P07).  This deep pointing highlighted the presence
of three other sources around S2, of luminosity and spectral shape
consistent with those of low mass X-ray binaries; only one is
consistent with being pointlike, though, the other two (S1 and S4) are
extended (Tab.~\ref{s1-s4}), and could be due to the superposition of
few point sources and/or truly diffuse emission. In particular, the
morphology of S1 resembles a jet-like feature (see also Fabbiano et
al. 2004).

In this paper we report the results of an observational campaign aimed
at complementing the $Chandra$ results, to better constrain the nature
of the processes taking place in the vicinity of the MBH in
NGC821. New data presented in this paper include sensitive VLA and
$Spitzer$ IRAC observations; we also analyzed archival $Hubble$
$Space$ $Telescope$ ($HST$) observations. With this multi-wavelength
data set we investigate: circumnuclear starformation, which may result
from accretion (Tan \& Blackman 2005); radiatively inefficient
accretion, possibly coupled to a compact or resolved nuclear jet
(e.g., Narayan 2005); and highly obscured scaled-down AGN
emission. The paper is organized as follows: in Section~\ref{endistr}
we summarize the main $Chandra$ results concerning the nuclear
sources, we present the results of the new VLA and $Spitzer$
observations, of the archival $HST$ optical data, and previous NICMOS
results; in Section~\ref{quie} we discuss possible scenarios for
accretion around this MBH; in Section~\ref{concl} we
summarize our conclusions.

\section{The emission properties of the nuclear region}\label{endistr}

\begin{figure}
\plotone{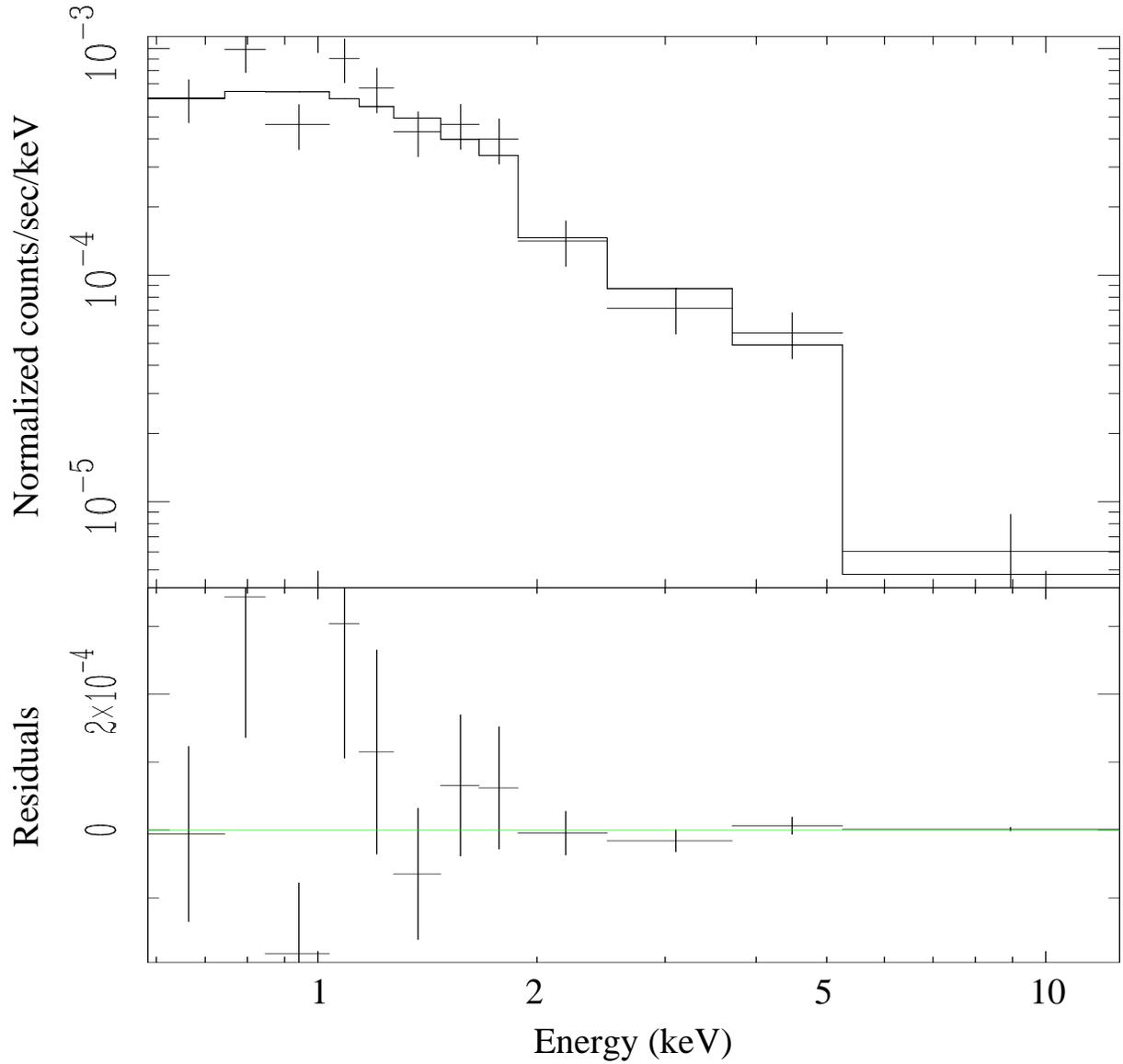}
\caption{The $Chandra$ ACIS-S spectrum of the source S2 detected at the
center of NGC~821 with the deep pointing (P07). Upper panel:
data and best fit power law model; lower panel:
the residuals from the best fit (Tab.~\ref{tabspec})}\label{s2spec}
\end{figure}

The main outcomes of the deep $Chandra$ pointing concerning the
properties of the nuclear source S2 are summarized in
Tab.~\ref{tabspec}.  We also show the source spectrum
(Fig.~\ref{s2spec}) and the results of a spectral fit including
fluorescent line emission from cold iron at 6.4 keV
(Tab.~\ref{tabspec}), which produced only an upper limit on the
equivalent width of the line.  Tab.~\ref{s1-s4} summarizes the
properties of the other two extended sources (S1 and S4) detected by
$Chandra$ in the central galactic region. The position and extent of
the sources S1, S2 and S4 are also shown in Fig.~\ref{radioim}.

Below we present new VLA observations in Sect.~\ref{vla}; the analysis
of IR data from a $Spitzer$ program to observe a sample of radiatively
quiescent MBHs (PI: Fabbiano) in Sect.~\ref{spiz}, together with a
previous NICMOS measurement; the reanalysis of archival $HST$ WFPC2
data taken with the F555W and F814W filters, and of our previous
R-band INT images in Sect.~\ref{opt}. Table~\ref{tablesed} summarizes
the results and Fig.~\ref{sed} shows them.

\subsection{Radio observations}\label{vla}

We observed NGC821 with the VLA of the NRAO\footnote{The National
Radio Astronomy Observatory is a facility of the National Science
Foundation operated under cooperative agreement by Associated
Universities, Inc.} at 1.4 GHz in the most extended (A) configuration
on 2004 December 31, and at 4.8 and 8.4 GHz in the second most compact
(C) configuration on 2005 July 21.  We used a 200 MHz instantaneous
continuum bandwidth, observing 3C48 for flux density calibration and
J0203+1134 for phase calibration, using standard techniques and the
AIPS data reduction package.  NGC821 lies $1.5^\circ$ from the phase
calibrator.  Empirical limits on systematic position uncertainties
were estimated from snapshots images of three quasars (J0204+1514,
J0149+0556, and J0242+1101) that lie within $10^\circ$ of the phase
calibrator.  The flux density scale was accurate to $<3 \%$ and
absolute position measurements were accurate to $0\rlap{.}''1$.  We
mapped the entire primary beam of the VLA with 3-D deconvolution to
obtain noise levels ($1\sigma$) of 20 $\mu$Jy, 20 $\mu$Jy, and 17
$\mu$Jy at 1.4, 4.8, and 8.4 GHz respectively, for effective total
integrations of 2.0--2.2 hours at each frequency.  We adopted variance
weighting of ({\it u,v})-data with some downweighting of short
baselines (AIPS parameter ROBUST=0).  Beam half-power sizes are given
in Tab.~\ref{tablesed}.

\begin{figure*}
\plotone{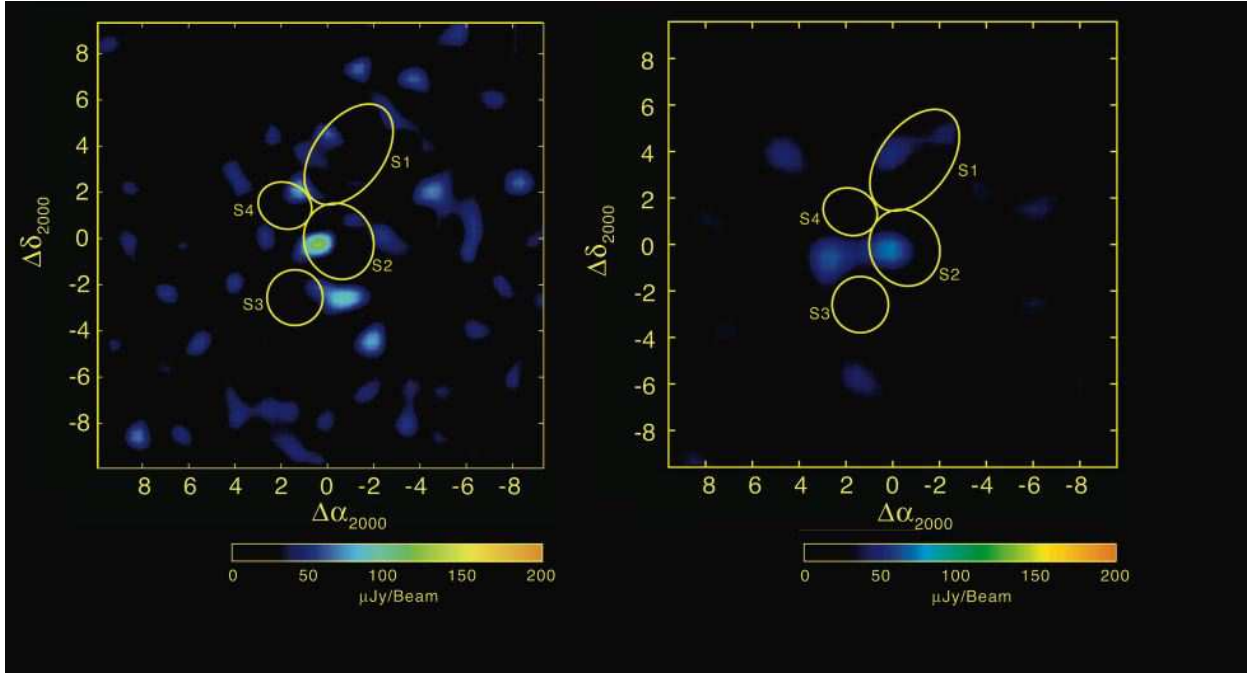}
\hskip 1truecm
\caption{Radio image of the central $17''$ ($\sim 2$\,kpc) of NGC821
at 1.4 GHz (left) and 8.4 GHz (right). The position and sizes of the
ellipses corresponding to the $Chandra$ sources S1--S4 (P07) are 
superposed.  The beam size was $1\farcs3\times 1\farcs3$ at 1.4 GHz and
$2\farcs7\times 2\farcs 2$ at 8.4 GHz (half-power full width).  The
coordinate origin is $02^h08^m21\rlap{.}^s140$, $+10^\circ
59'41\rlap{.}''70$ (J2000). } \label{radioim} 
\end{figure*}

We detected a radio counterpart to source S2 at $\alpha_{2000} =02^h
\, 08^m \, 21\rlap{.}''174\pm0.007$, $\delta_{2000}=10^\circ 59'
41\rlap{.}''4\pm0\rlap{.}''1$ at 1.4 GHz (Fig.~\ref{radioim},
Tab.~\ref{tabspec}; Sect. 4.2 of P07 discusses 
the coincidence of the position of S2 with the optical center in 
$HST$ WFPC2 images), with a flux density of 127 $\mu$Jy
(6$\sigma$).  Emission was marginally detected at 8.4 GHz (71 $\mu$Jy;
4.2$\sigma$), peaking at the same position to within the measurement
uncertainty. The noise level at 8.4 GHz was about one and a half
order of magnitude lower than the earlier upper limit obtained at 15
GHz by Nagar et al. (2005).  A marginal detection at 4.8 GHz was also
possible (80$\mu$Jy; 4.0$\sigma$); in this case the prospective peak lies
$\sim 2''$ (or $1\over2$ beamwidth) to the south. 

The three flux densities correspond to a power law with spectral index
$\alpha = 0.33\pm0.04$ ($S_\nu\propto \nu^{-\alpha}$).  The estimated
spectral index may be viewed with some caution because it depends on
measurements made with very different beamwidths.  However, we have
studied images at 1.4 GHz made with different weighting schemes (i.e.,
Natural and Uniform) and we did not find a significant change in peak
signal to noise ratio at the location of S2, for beam sizes of $1\farcs2$
to $1\farcs6$. At this and other frequencies, there is no indication that 
we might be partially resolving the emission for S2; however, 
multi-configuration data would be needed to properly address this question.

No sources were detected at the location of the other $Chandra$ sources S1, S3
and S4 in the central galactic region (Sect.~4 in P07; Fig.~\ref{radioim}).
Therefore, we can place $4\sigma$ upper limits on their emissions of 80$\mu$Jy,
80$\mu$Jy and 68$\mu$Jy respectively at 1.4, 4.8 and 8.4 GHz.

\subsection{Infrared observations}\label{spiz}

NGC821 was observed with the Infrared Array Camera (IRAC, Fazio et
al. 2004) on board the $Spitzer$ Space Telescope (Werner et al. 2004)
on August 21, 2005 for a total of 725.5 s (Program ID 20371). After
the standard $Spitzer$ IRAC
processing\footnote{http://ssc.spitzer.caltech.edu/irac/dh/PDD$\_$v1.pdf},
the four IRAC channels resulted in images at 3.6, 4.5, 5.8 and 8 $\mu
m$; the central region of NGC821 was detected in each of them. We
based our analysis on the IRAC instrument performance as given in the
Infrared Array Camera Data
Handbook\footnote{http://ssc.spitzer.caltech.edu/irac/dh/iracdatahandbook3.0.pdf}.

Using DS9 and $funtools$\footnote{hea-www.harvard.edu/RD/funtools/ },
we extracted counts from circular regions of approximately
4 square arcseconds area ($\sim 1\farcs1$ radius) centered on the peak of
the emission. In every case the position of this peak, with the $Spitzer$ WCS,
falls in-between the optical center and the $Chandra$ position for S2
(see Tab.~\ref{tabspec}); both of these are
well within the count extraction radius. This position ($Spitzer$ WCS)
is at RA=$02^h$ $08^m$$21^s$\hskip-0.1truecm.11, Dec=$+10^{\circ}$
$59^\prime$ $42\farcs0$, although given the pixellation of the data,
the peak count pixel may be slightly displaced.

The background was estimated both from a surrounding annulus, to
attempt a rough subtraction of stellar light from the center-most
emission, and from an off-source circle. The fluxes were all normalized to
an area of 4 square arcseconds (that of circles of  $1\farcs1$ radius).  The
results are shown in Table~\ref{spizflu}, for both choices of the
background.

We also list in Tab.~\ref{tablesed} the upper limit on any
unresolved nuclear emission in the H-band (centered at 1.6$\mu m$)
derived from NICMOS data (Ravindranath et al. 2001).
This limit refers to the NIC2 camera and corresponds to an
intrinsic Gaussian of FWHM $\la 0.5$ pixel.

\subsection{Optical observations}\label{opt}

We carried out aperture photometry on the archival $HST$ WFPC2 images
taken with the F555W and F814W filters (Lauer et al. 2005), following
standard procedures described in the WFPC2 handbook\footnote{See
http://www.stsci.edu/instruments/wfpc2/Wfpc2$\_$hand/wfpc2$\_$tutorial$\_$v3.0.pdf.}.
The instrumental F555W and F814W magnitudes were converted to standard
V and I magnitudes using the most updated version\footnote{See
http://purcell.as.arizona.edu/wfpc2$\_$calib/.} of the color
coefficients provided by A. Dolphin. In order to compare
the optical emission with that estimated from $Spitzer$ IRAC
observations, it was calculated for a circle of $1\farcs1$ radius, the
same area used to extract the $Spitzer$ fluxes (Sect.~\ref{spiz}).  The
optical brightness inside this area is $V_0 = 15.0 \pm 0.1$ mag, and
$I_0 = 13.7 \pm 0.1$, after correcting for a Galactic extinction
$E(B-V) = 0.11$ (Schlegel et al. 1998).  The R-band magnitude from the
same area was also calculated, by using INT images (Graham et
al. 2001) and the corresponding surface brightness profile derived by
Soria et al. (2006b); it results into $R_0 = 14.3 \pm 0.1$ in the
Cousins system.  

In order to constrain the emission coming from the MBH, from the
deconvolved F555W and V--I profiles derived by Lauer et al. (2005) we
also calculated the V and I luminosities within a circle of
$0\farcs046$ radius (the innermost radius at which deconvolution can
give accurate results according to Lauer et al.).  These luminosities
are upper limits to the MBH emission, since both our analysis and that
of Lauer et al. (2005) reveal the absence of an optical "nucleus",
that is a compact light source rising above the surface brightness
profile at small radii when extrapolated inward.  In general these
nuclei, which have been found in a large fraction of early-type
galaxies from $HST$ images (see also Ravindranath et al. 2001), are
bluer than the background starlight and could be nuclear star clusters
or low luminosity AGNs.

\section{Discussion}\label{quie}

The deep $Chandra$ image of NGC821 revealed an extended, hard source
at the position of the galactic center (P07); our associated VLA
observations have led to the discovery of a 1.4 GHz source at the same
position (Sect.~\ref{vla}); pointlike nuclear emission is not detected in the
optical and the IR (Sects.~\ref{spiz} and \ref{opt}).  In the 0.3--8
keV band, the $3\sigma$ upper limit to any pointlike emission
associated with the MBH is $2.8 \times 10^{38} ~\rm erg~s^{-1}$, and
the $\nu L_{\nu}$ at 1.4 GHz is just $1.2\times 10^{35}$ erg
s$^{-1}$. Therefore the MBH of NGC821 is one of the radiatively
quietest MBH known, with $L_X/L_{Edd}<2.5\times 10^{-8}$.  Most MBHs
are radiatively quiescent in the local universe, or show very low
activity levels (Sect.~\ref{intro}), therefore our deep look at the
nucleus of NGC821 has a bearing on a very common state of MBHs.  In
the following we use our observational results to constrain the
process of accretion in this nucleus.

\subsection{Why not a dead MBH?}\label{dead}

Given the general lack of fuel available for accretion observed at all
wavelengths, including the X-rays (Sect.~\ref{intro}), is a truly dead
MBH what we should expect at the center of NGC821?  In fact, an aging
stellar population continuously returns gas to the ISM, via its
stellar mass losses (Ciotti et al. 1991, David et al. 1991), and the
circumnuclear region should be replenished with this fuel. Indeed,
hydrodynamical simulations specific for NGC821 (P07) demonstrate that
this fuel should be present.  While the bulk of the hot gas is
expelled out in a wind (consistent with the lack of detection of hot
gas), the stellar mass losses are accreting within a
very small region of $\sim 25 $ pc from the nucleus; at the innermost
radius resolved by the simulations (10 pc), the mass accretion rate is
$\dot M_{in}\approx $few$\times 10^{-5}M_{\odot}$
yr$^{-1}$. Notwithstanding the limitations of the simulations, and the
uncertainties associated with the observables used as input parameters, 
the presence of this small inflowing region was considered
a robust result (see P07); similarly, a range of values for $\dot
M_{in}=(2-7)\times 10^{-5}M_{\odot}$ yr$^{-1}$ was considered reliable
(with the caveat that the true, final accretion rate on the MBH could
be somewhat higher; P07). $\dot M_{in}$ should produce a luminosity
$L_{acc}\sim \epsilon \dot M c^2 \sim $(1--4)$\times 10^{41}$ erg
s$^{-1}$ if, closer in to the MBH, it ends up in a standard accretion
disc with a radiative efficiency $\epsilon \sim 0.1$ as assumed for
classical AGNs (Shakura \& Sunyaev 1973).  This would be a highly
sub-Eddington AGN, though, since $L_{acc}/L_{Edd} \sim $few$\times
10^{-5}$. For a standard spectral energy distribution (SED) of AGNs
(Elvis et al. 1994), the upper limit on a pointlike nuclear X-ray
emission (Tab.~\ref{tabspec}) corresponds to a bolometric upper limit
of $L_{bol}< 4\times 10^{39}$ erg s$^{-1}$, that is already 30 to 100 times
less than $L_{acc}$.

Below we explore scenarios for the evolution of the accreting
material, that would explain why the nucleus is not observed at a
luminosity as high as $L_{acc}$. We consider, in
turn, the presence of angular momentum in the accreting material
(Sect.~\ref{ang}), a reduction of the accretion rate due to
starformation in a circumnuclear disc (Sect.~\ref{bc03}), accretion at
low radiative efficiency (Sect.~\ref{effic}), the presence of a
jet/outflow, whose kinetic power possibly accounts for a fraction of
$\dot M_{in}c^2$ (Sect.~\ref{aneta}), a
transition in the accretion mode due to disk instabilities
(Sect.~\ref{instab}) and finally obscuration of the nuclear radiation
produced by a scaled-down AGN (Sect.~\ref{osc}).

Note that the mass accretion rate of galactic nuclei has been often
estimated (e.g., Di Matteo et al. 2003) assuming spherically symmetric
accretion from a nonrotating polytropic gas with given density and
temperature at infinity (Bondi 1952).  Moscibrodzka (2006) calculated
model spectra emitted by such a Bondi flow for a sample of low
luminosity AGNs, including NGC821. The adopted value of gas density
was larger than constrained now thanks to the deep $Chandra$ pointing
(P07); even so, the predicted X-ray emission was $< 2.6 \times
10^{33}$~ergs~s$^{-1}$, that is $\sim 5$ orders of magnitude below the
limit on pointlike nuclear emission (Tab.~\ref{tabspec}).  However,
the Bondi accretion is a severe mathematical approximation of the
accretion process (see also P07); for example, some initial angular
momentum of the flow captured by the MBH, and the inclusion of
viscosity, may result in  a higher X-ray
luminosity.  Radiation models applied to global MHD simulations of low
angular momentum accretion flows are being developed and will be
applied to this nucleus (Moscibrodzka et al., in preparation; see also
Balbus \& Hawley 2002).

\subsection{Angular momentum at large radii}\label{ang}

The inclusion of even slow rotational motion of the gas at large radii
can significantly reduce the rate at which mass is captured into the
accretion flow, compared to the non-rotating case (Proga \& Begelman
2003). The motivation for the Proga \& Begelman's study was to make
less severe the problem that MBHs in the local universe are much
dimmer than expected for a mass accretion rate estimated at large
radii via the Bondi (1952) formula (Sect.~\ref{dead}) and ending up in
a standard disc (e.g., Fabian \& Canizares 1988, Pellegrini
2005a). While the Bondi formula assumes spherically symmetric
accretion from a nonrotating gas under the sole influence of the
central gravity, Proga \& Begelman (2003) studied the evolution of the
flow with a small angular momentum, including regions beyond the
domination of the MBH gravity.  They found that the mass supply rate
to the MBH can be smaller by up to $\sim 2$ orders of magnitude with
respect to the nonrotating case.  In NGC821 the stellar kinematics
becomes disc-like within the central $\sim 10^{\prime\prime}$ (Scorza
\& Bender 1995, Emsellem et al.  2004), therefore the stellar mass
losses close to the galactic center are likely to have angular momentum,
which would lead to a reduction of $\dot M_{in}$.

Whether the expected luminosity could become consistent with the
observed upper limits is however more questionable, since it is not
known what is the final, net effect of the inclusion of angular
momentum at large radii on the resulting luminosity of the flow closer
to the MBH. It is probably not plausible that gas inflowing at the
rate shown by the simulations (that is equal to $\dot M_{in}$ of
Sect.~\ref{dead} at present and was larger in the past, for a total
integrated mass of few$\times 10^6~M_{\odot}$ over the past $\sim 10$ Gyrs; 
P07) continues to
accumulate for many Gyrs in the nuclear region without accreting and
without becoming observable. However, in order for the gas to be able to
accrete, angular momentum must be transferred outward through
viscosity, and this may actually increase the luminosity of the
flow. MHD simulations with rotation and cooling by radiation included
are needed to really state what happens to the luminosity of the flow
(see also Sect.~\ref{dead}).

\subsection{Circumnuclear starformation}\label{bc03}

\begin{figure*}[ht]
\vskip -6truecm
\plotone{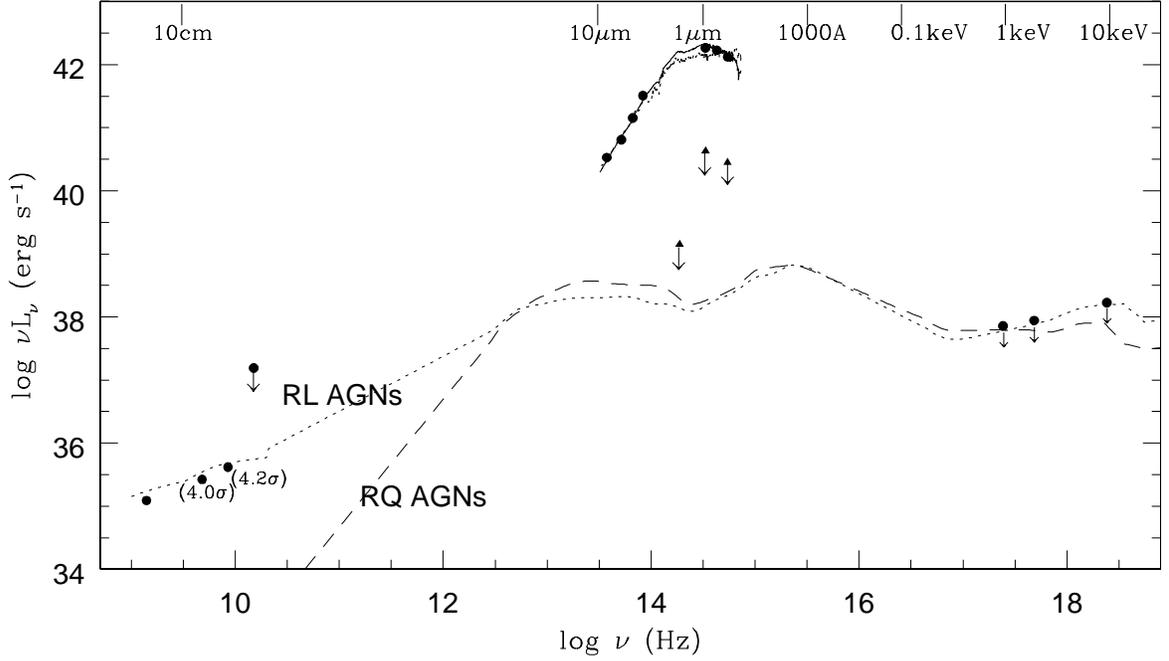}
\vskip -1truecm
\caption{The spectral energy distribution of the nuclear emission of
NGC821 (data in Tab.~\ref{tablesed}; see also Sect.~\ref{endistr}).
The radio detection at 1.4 GHz, and the $4.0\sigma$ and $4.2\sigma$
detections at 4.8 and 8.4 GHz, come from our VLA observations; the 15
GHz upper limit is from Nagar et al. (2005, Sect.~\ref{vla}). The
$Spitzer$ IRAC measurements from 8 to 3.6$\mu m$ derive from our
$Spitzer$ observations and refer to a circle of $1\farcs1$ radius
(Sect.~\ref{spiz}) for the field background; the upper limit at
1.6$\mu m$ (shown with a triangle) is from NICMOS data (Ravindranath et al. 
2001) for a circle of $0\farcs02$ radius. The I, R and V measurements refer 
to the same extraction area used for the $Spitzer$ data, and derive 
from our analysis of the $HST$ WFPC2 F814W and F555W image, and of 
an INT image (Soria et al. 2006b). From
these two $HST$ images, an upper limit (shown with triangles) to the 
nuclear emission
(i.e., the luminosity enclosed within $0\farcs046$) has 
been derived from the deconvolved profile of Lauer et al. (2005, see
Sect.~\ref{opt}). The 1, 2, 10 keV upper limits refer to a pointlike
emission possibly associated with the MBH, from  
$Chandra$ data (Tab.~\ref{tabspec}; P07). Dotted and dashed
lines show the median distribution observed for low redshift radio
loud and radio quiet AGNs (Elvis et al. 1994), rescaled to match the
X-ray upper limits. The solid line gives the best fit SED of Bruzual
\& Charlot (2003) to the $Spitzer$ and I, R, V data (see
Sect.~\ref{bc03}); also shown (dotted line) is the best fit SED for
higher abundance and younger age.}\label{sed} 
\vskip -4truecm
\end{figure*}

Condensation into stars in a gravitationally unstable disc can
prevent a significant fraction of the accreting gas from reaching the
MBH (Tan \& Blackman 2005), resulting into a much lower true accretion
rate; this starformation should produce H$\alpha$ and IR emission.  As
reported in Sect.~\ref{intro}, the optical spectrum of NGC821 shows
only absorption lines; the $3\sigma$ upper limit on its H$\alpha$
luminosity is $L_{H\alpha}<1.46\times 10^{38}$ erg s$^{-1}$ (Ho et
al. 2003, rescaled for the distance in Tab.~\ref{mainlog}). From the
Kennicutt (1998) relation between starformation rate and
$L_{H\alpha}$, the resulting starformation rate is $<10^{-3}
M_{\odot}$ yr$^{-1}$, much larger than $\dot M_{in}$ given by the
simulations (Sect.~\ref{dead}).  Also the observed luminosity in the
IRAC bands (Tab.~\ref{tablesed}, Fig.~\ref{sed}) is at least one order
of magnitude higher than expected in the full IR band from
starformation at the $\dot M_{in}$ rate (Kennicutt 1998).  However,
the aperture used to derive the IRAC luminosities includes also
emission from the normal, old stellar population, that has to be 
accounted for.

The analysis of the SED of NGC821 (Fig.~\ref{sed}) shows that it 
can be explained just by the normal, old galactic stellar population,
without requiring  ongoing starformation. To check this,
we used the matched aperture photometry (i.e., all for
$1\farcs1$ aperture radius) from the $Spitzer$ to the V filter data
points (Tab.~\ref{tablesed}) and compared it with SED templates for
stellar populations of different ages and metallicities, for the
Kroupa (2001) IMF (Bruzual \& Charlot 2003); we also allowed for an
exponentially declining starformation rate with time scale $\tau$. The
fitting procedure described by Bolzonella et al. (2003)\footnote{This
fitting can be made with the public code $hyperz$, available on the
web at http://webast.ast.obs-mip.fr/hyperz and built by Bolzonella et
al. (2003).} was applied.  The observed data points are very well
fitted by an age of 11.0 Gyrs and a metallicity of 0.25 solar, with a
short time scale ($\tau=0.1$ Gyr) and without current star formation
(solid line in Fig.~\ref{sed}).  These properties match closely those
obtained from spatially resolved optical spectroscopy for the bulk of
the stellar population (Proctor et al. 2005).  The magnitudes and
fluxes derived here have been corrected for extinction
(Sect.~\ref{endistr}); to check whether a residual extinction is left
in the data, we allowed for free extinction in the fit, but the best
fit extinction turned out to be consistent with zero.  We can conclude
that the optical and mid-IR emission within a central projected circle
of $1\farcs1$ radius is dominated by the normal stellar population of
the galaxy.

From their analysis based on Lick absorption-line indices, Proctor et
al. (2005) also found at the very center of NGC821 (within a central
radius of $1^{\prime\prime}$) an emission weighted abundance of $\sim
3$ times solar, and evidence for a burst of starformation between 1
and 4 Gyrs ago, that involved $<10$\% of the galaxy's mass.  Therefore
we fitted the observed data points also with a metallicity of 2.5
solar (the largest value available for the templates); this 
resulted in an age of $\sim 1$ Gyr, but the fit was not as good as the
previous one (dotted line in Fig.~\ref{sed}).  The optical points
considered here contain less information than the indices used by
Proctor et al. (2005); it is also likely that the true best fit SED is
a combination of templates with different ages and abundances, but a
fit with composite SEDs is beyond the scope of this work.  For the
purposes of the present investigation, the conclusion here is that
current starformation is not required to explain the SED; however,
$\dot M_{in}$ is so low that the available optical and IR data cannot
constrain whether a part (or even the total) of it goes into star
formation.

\subsection{Low radiative efficiency (plus a compact, nuclear jet)}\label{effic}

It has been suggested that high-luminosity and low-luminosity AGNs
host two different accretion modes: a radiatively efficient
disk-dominated one and a radiatively inefficient one, which is
expected to be coupled to strong outflows (see Narayan 2005 for a
recent review). The SED of a few low-luminosity AGNs indeed has been
modelled with truncated optically thick disks with inner radiatively
inefficient flows (e.g., M81 and NGC4579, Quataert et al. 1999;
NGC3998, Ptak et al. 2004; NGC1097, Nemmen et al. 2006). In the
radiatively inefficient mode, the flow produces the X-ray emission,
while the radio emission comes from the base of a nuclear jet, if
present (e.g., Falcke et al. 2004, K\"ording et al. 2006). In fact,
observations seem to suggest that low-luminosity AGNs are "radio-loud"
(Ho 2002, Terashima \& Wilson 2003, Wu \& Cao 2005); moreover, often
their radio power and sometimes also their radio spectrum require a
separate, compact jet component (Anderson et al. 2004; Nagar et
al. 2005). On the theoretical side, MHD simulations show that at low
accretion rates winds and jets naturally develop (e.g., Stone et
al. 1999, Hawley \& Balbus 2002, Igumenshchev et al. 2003, De Villiers
et al. 2003), since the magnetic and kinetic energy release in the
flow are enough to support an outflow or a jet (see also Blandford \&
Begelman 2004).

In the context of low radiative efficiency accretion, the relationship 
between the black hole mass, the
core radio luminosity at 5 GHz and the 2--10 keV emission has been
investigated for a large sample of accretion-powered sources, going
from X-ray binaries to mostly low-luminosity AGNs by Merloni et
al. (2003).  They found the sources to lie on a plane described by log
$L_R=0.60$ log $L_X +$ 0.78 log $M_{BH} + $7.33 (with a scatter of
$\sigma_R=0.88$), which was attributed to the presence of a
radiatively inefficient accretion flow $+$ jet system, that can exist
with similar properties at different scales.

\begin{figure*}[ht]
\vskip -6truecm
\plotone{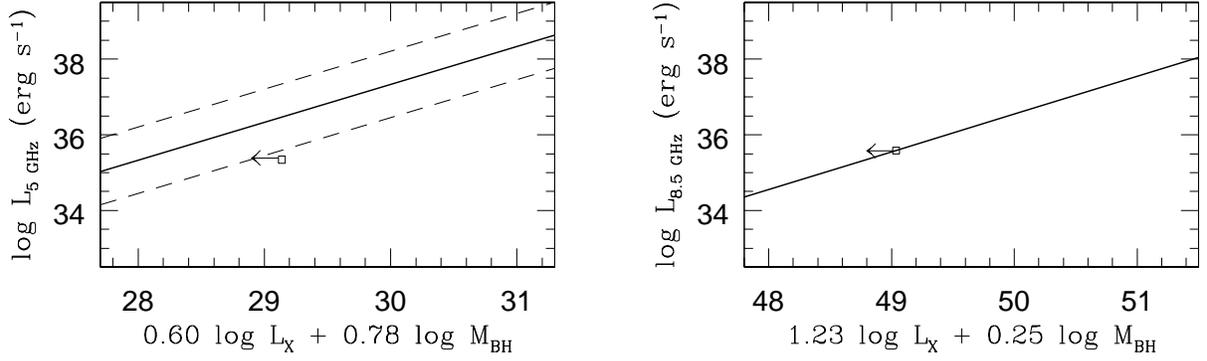}
\caption{Location of the nucleus of NGC821 with respect to the fundamental
plane of black hole activity. Left: the plane of Merloni et
al. (2003) is indicated with a solid line,
together with the scatter of $\sigma_{log\,R}=0.88$ of
observed sources (dashed lines); right: the plane of Yuan \& Cui
(2005). See also Sect.~\ref{effic}.}\label{fp}
\end{figure*}

Low radiative efficiency accretion is compatible with the very low
value of $L_X/L_{Edd}<2.5\times 10^{-8}$ for the MBH of NGC821
(Sect.~\ref{quie}), and was already considered as a viable solution
for its low emission by Pellegrini (2005a) and Soria et al. (2006a,b).
Given our new radio observation, we examine here the position of this MBH with
respect to the plane of Merloni et al. (2003). The 2--10 keV upper
limit on nuclear pointlike emission is $1.8\times 10^{38}$ erg
s$^{-1}$ (Tab.~\ref{tabspec}), and the nuclear 5 GHz luminosity is
$2.7\times 10^{35}$ erg s$^{-1}$ (a $4.0\sigma$ detection, though;
Sect.~\ref{vla}); the resulting location is shown in Fig.~\ref{fp}. If
the 5 GHz detection is real, and the nuclear X-ray emission is smaller
by a factor of $\ga 2$ than the upper limit, then NGC821 is consistent
with the plane, and it may host a radiatively inefficient accretion
flow $+$ jet system.  If instead the X-ray emission is close to that
of the whole S2 source, or even higher (see Sect.~\ref{osc}), then
this nucleus lies well below the predictions of the plane. In this
case, its SED could be consistent with that of a classical, bright AGN
(see Sect.~\ref{osc} below).

The sample of Merloni et al. includes mostly objects of relatively
high $L_X/L_{Edd}$.  Below a critical $L_X/L_{Edd} \sim 10^{-5}$, the
jet emission can become dominant also in the X-ray domain, and the
radio--X-ray correlation should turn steeper (Yuan \& Cui 2005).  The
slope of the fundamental plane should then be different from when the
X-ray emission is dominated by the accretion flow (see also Wang et
al. 2006), and the fundamental plane in this regime becomes log
$L_R=1.23$ log $L_X +$ 0.25 log$M_{BH} - $13.45 (with $L_R$ at 8.5
GHz, Yuan \& Cui 2005). This relation predicts a lower radio
luminosity for given X-ray emission and therefore the nucleus of
NGC821 is located close to such a plane (Fig.~\ref{fp}) provided that
the radio emission is close to the $4.2\sigma$ detection at 8.4 GHz
and the 2--10 keV emission close to the $3\sigma$ upper limit.

In conclusion, a radiatively inefficient flow coupled to a nuclear jet
is a viable scenario for the accretion process. Also the observed
spectral shape of the nuclear source in the radio is consistent with
this interpretation.  The flux density of a jet with optically thick
core/nozzle emission is approximately flat ($S_{\nu}\propto \nu
^{-\alpha}$, with $\alpha\sim 0$) up to high radio frequencies (Zensus
1997, Falcke et al. 2004).  Within the uncertainties due to the
marginal detections at 4.8 and 8.4 GHz and the unmatched beamwidths at
the different frequencies (Sect.~\ref{vla}), the radio spectral shape
of the nuclear source is relatively flat, and is therefore consistent
with being produced by the base of a jet. This jet could be confined
within the inner few arcsecs (see Tab.~\ref{sed} for the beam sizes),
because it is disrupted or frustrated within it; nuclei with
parsec-scale jets have already been found in elliptical galaxies
with the VLBI (e.g., Nagar et al. 2005).

In the next Section we examine whether a jet/outflow can be present on
a larger scale (outside S2).

\subsection{A resolved jet?}\label{aneta}

The elongated X-ray sources S1 and S4 in the central galactic region
(P07, Fig.~\ref{radioim}) may be due to an outflow or a jet; in the
radio, only S2 is detected, but at the location of sources S1 and S4
we could place $4\sigma$ upper limits (Sect.~\ref{vla}).  We use here
the observed X-ray to radio flux ratios for typical jets to
investigate whether S1 and S4 can be due to a jet (note however that
the properties of jet emission depend on the internal jet structure as
well as the environment in which the jet propagates). Kataoka \&
Stawarz (2005) compiled a sample of X-ray jets with $Chandra$
data, observed outside the host galaxy of
quasars and AGN. The X-ray luminosity usually dominates over the
radio for the jet/knots emission, while for lobes and hot spots the
radio luminosity is comparable to or exceeds the X-rays. In any case,
in the Kataoka \& Stawarz (2005) sample the X-ray to radio
L$_{1keV}$/L$_{5GHz}$ monochromatic luminosities ratio does not exceed
100. This same ratio is $>552$ for the source S1, and $>375$ for S4,
assuming that the entire X-ray luminosity of each source is due to a
jet, and the $4\sigma$ upper limits at 5 GHz (Sect.~\ref{vla}).
However, if the observed X-ray luminosity is contributed by several
blended knots then the limits quoted above are not meaningful.
The case for unresolved knots in
sources S1 and S4 is better investigated with more nearby objects.  One
example of a nearby jet moving within the interstellar medium of a
galaxy is that of Centaurus A, with many resolved X-ray knots
(Hardcastle et al. 2003, Kataoka et al. 2006). Cen A is at 3.4~Mpc
distance, so that $1^{\prime\prime}$ is equivalent to 17~pc.  The size
of source S1 (Tab.~\ref{s1-s4}) corresponds to $\sim 600$ pc, and it
is feasible for a jet to create many knots on that distance, given the
large number of knots observed in the Cen~A jet. Also,
for a few knots in the Cen~A jet the X-ray to radio
luminosity ratio exceeds 100, with one case of $>500$ (Hardcastle et
al. 2003). The lowest detected 5~GHz luminosity of a radio knot in the
Cen~A jet is 1.4$\times10^{35}$ erg~s$^{-1}$, which is below our
$4\sigma$ upper limit 
of 2.7$\times10^{35}$ erg~s$^{-1}$ at 5 GHz.

The available radio and X-ray data, although very deep, do not provide
therefore a definite conclusion on the jet presence outside the S2
region.  

Note that our current data do not resolve the nuclear S2 source, whose
X-ray and radio luminosities are measured for a region of $\sim 200$ pc
radius, and could also be related to a small jet. However, the radio
spectrum of S2 ($\alpha =0.33$, Sect.~\ref{vla}) does not have the
typical slope observed for extragalactic jets or hot spots ($\alpha
\sim 0.6-0.8$, Kataoka \& Stawarz 2005), but is instead more similar
to that of radio-loud AGNs (Elvis et al. 1994), and
therefore more likely produced by the base of a jet.

If present, the launch of a nuclear jet/outflow could account for a
non-negligible fraction (of the order of a few percent) of the energy
associated with the rest mass of the material being accreted (i.e., of
$\dot M_{in}c^2$), as shown to be the case for M87 (Di Matteo et
al. 2003), IC1459 (Fabbiano et al. 2003) and IC4296 (Pellegrini et
al. 2003) and recently in a few other hot gas rich sources (Allen et
al. 2006).  This may be a characteristic of the late stages of galaxy
evolution, when accretion onto the MBH falls below the Eddington rate
(e.g., Churazov et al. 2005). Unfortunately, radio synchrotron
emission traces poorly the true jet power (Owen et al. 2000, B\^irzan
et al. 2004), with the ratio of the mechanical (kinetic) luminosity to
the 1.4 GHz synchrotron luminosity ranging between a few and roughly a
few thousand for luminous radio sources, and up to several thousand
for weaker sources. Notwithstanding these uncertainties, we can estimate that
the jet kinetic power could account for just a few$\times 10^{-4}$ of
$\dot M_{in}c^2$, for the nuclear radio source coincident with S2 and
detected at 1.4 GHz, and at most this fraction for the S1 and S4
sources with upper limits at 1.4 GHz.  An injection of energy from the
nucleus may have heated the surroundings, so that accretion has been
recently stopped or lowered to values below the $\dot M_{in}$
estimated in the numerical simulations of P07 that do not include
feedback (Di Matteo et al.  2003; Ciotti \& Ostriker 2007).

\subsection{Disk instability}\label{instab}

The mode of accretion may switch to a quiescent state due to the
change in internal structure of the accretion flow produced by disk
instabilities (Siemiginowska et al.  1996, Janiuk et al. 2004), with
the timescale in the low quiescent state longer than that in the
active state. Then the accretion flow can be described by a
hybrid model, where an outer torus or cold disk accumulates accreting
matter at the steady accretion rate, while the inner torus/disk has a
low accretion rate and is radiatively inefficient. The transition
radius between the two regions is set by the hydrogen ionization
instability and depends on the accretion conditions onto the outer
torus.

For the MBH mass of NGC821 (Tab.~\ref{mainlog}) the gravitational
radius is located at $R_g=2.5 \times 10^{13}$ cm. Assuming an
accretion rate of $2\times 10^{-5} M_{\odot}$~yr$^{-1}$
(Sect.~\ref{dead}), the location of the ionization zone is at $\sim
100 R_g$ (Janiuk et al. 2004). In the hybrid model we expect the outer
disk to be cold and to emit a thermal spectrum with a total
disk luminosity for these parameters of about $2.5\times 10^{37}$
ergs~sec$^{-1}$, that will be radiated mainly in the optical and near
IR bands. This value is well within the observed limits
(Tab.~\ref{tablesed} and Fig.~\ref{sed}).  The X-ray luminosity will
come from the inner quasi-spherical accretion flow and
depend on the density and temperature of the plasma; its value
needs to be self-consistently calculated, which is beyond the scope of
this work and will be addressed by Moscibrodzka et al. (in preparation).

\subsection{A standard AGN (but very obscured)}\label{osc}

We examine here the possibility that the nucleus of NGC821 is a normal
AGN working at a very sub-Eddington rate of $L_{acc}/L_{Edd}\sim$few$\times
10^{-5}$, with $L_{acc}\sim (1-4)\times 10^{41}$ erg s$^{-1}$
estimated for a standard disc as described in Sect.~\ref{dead}, and 
test this hypothesis using the whole observed SED (Fig.~\ref{sed}).  
$L_{acc}$ is 30--100 times larger than the upper limit on $L_{bol}$
of pointlike nuclear emission (for a bolometric correction of $L_{bol}\sim
20 L_{2-10\rm{keV}}$, Elvis et al. 1994).  We consider then the
possibility that the X-ray radiation is heavily absorbed, so that the
intrinsic $L_{bol}$ is comparable to $L_{acc}$.

A Compton thin AGN with $N_H\sim 10^{23-24}$ cm$^{-2}$ (e.g., Bassani
et al.  1999) can be excluded, since the X-ray spectrum of S2 does not
show the characteristic cut-off at low energies
(Tab.~\ref{tabspec}). A more interesting possibility is that S2 is a
Compton thick AGN, where the direct nuclear emission is not detectable
below $\sim 10$ keV because of an absorber of very large column
density ($N_H > 10^{24}$ cm$^{-2}$; e.g., Guainazzi et al. 2005).
However, some few percent of the intrinsic emission is often scattered
and reprocessed off some Compton thick 'mirror' visible both to us and
from the central continuum source, and may emerge below 10 keV with a
flatter photon index than the intrinsic X-ray continuum (i.e., with
$\Gamma \sim 1$), as for example in the Circinus galaxy (Matt et
al. 1999). The resulting X-ray spectrum also shows a very strong
(equivalent width $\ga 1$ keV) iron $K\alpha$ fluorescent line (as in
NGC1068, Levenson et al. 2006). S2 has quite a flat spectral shape
($\Gamma=1.49^{+0.14}_{-0.13}$), reminiscent of what is found for
Compton-thick sources; a 6.4 keV iron line is not seen in its X-ray
spectrum, but only with a $1\sigma$ upper limit on its equivalent
width of 2 keV (Tab.~\ref{tabspec}). Note also that extended emission
has been observed associated with some obscured Seyfert nuclei, up to
sizes of $\sim 1$ kpc (Elvis et al. 1990, Ogle et al. 2000), and is
understood as emission from gas photoionized by the central AGN, or
thermal emission from a hot collisionally ionized plasma that forms
the intercloud medium.  Therefore, this X-ray faint MBH may be an
extreme scaled-down version of such obscured AGNs, and the extended S2
source could be nuclear flux scattered and fluorescing off cold gas.
Evidence for Seyfert nuclei (including Compton-thick ones) being
scaled-down versions of more luminous AGNs down to 2--10 keV
luminosities of $\sim 10^{39}$ erg s$^{-1}$ has been provided recently
(Panessa et al. 2006, 2007), based on the significance of the correlation
between $L_{2-10{\rm keV}}$ and L(H$\alpha$) down to such low X-ray
luminosity values. This implies a similar proportion of X-ray
and UV ionizing radiation in high and low luminosity
nuclei, so that the shape of their SEDs may be similar in these
spectral regions.

If we assume that a typical 1\% of the intrinsic AGN flux is scattered
into our line of sight and observed as source S2 with a L(2--10
keV)$\sim 4\times 10^{38}$ erg s$^{-1}$ (Tab.~\ref{tabspec}), the
intrinsic $L_{bol}\sim 8\times 10^{41}$ erg s$^{-1}$ for a standard
AGN SED (Elvis et al. 1994), a luminosity that is consistent with
$L_{acc}$.  $L_{bol}$ could be even higher, since some nearby AGN show
$<1$\% of their X-ray flux scattered, perhaps due to the lack of an
appropriately positioned 'mirror' (NGC4051, Uttley et al. 2003).  An
intrinsic 2--10 keV luminosity as large as assumed here would place
the NGC821 nucleus way off the fundamental plane of Sect.~\ref{effic}
(increasing its x-axis value of 1.4 in Fig.~\ref{fp}), consistent with
the hypothesis of a standard disc rather than a radiatively
inefficient accretion flow.

How does this hypothesis of the nucleus of NGC821 being a scaled-down,
highly obscured AGN compare with the available observational
constraints?
 
The intrinsic SED would have the typical AGN shape of Elvis et
al. (1994) shown in Fig.~\ref{sed}, rescaled up by a factor of $\sim
200$ for a 1\% reflectance (since the average AGN SED in
Fig.~\ref{sed} is normalized to the upper limit on pointlike X-ray
emission in Tab.~\ref{tabspec}, that is $\sim 2$ times lower than the
whole X-ray luminosity of S2).  The upper limits in the V and I bands
allow for this shift, but the NICMOS F160W band limit lies just a
factor $\sim 10$ above both the radio-quiet and radio-loud SEDs before
rescaling.  Even though the dispersion in the SEDs of AGN is rather
large (even an order of magnitude in the X-ray to optical-UV ratio,
Risaliti \& Elvis 2004), scaling up the optical--IR SED of NGC821 by
only a factor of $\sim 10$ and the X-rays by a factor of $\sim 200$
would produce an $\alpha_{OX}$ index of $\la 1$, that is definitely
too low with respect to what typically observed (Risaliti \& Elvis
2004).  Therefore, a hidden AGN may have a $\sim 20$\% reflectance,
rather than 1\% as assumed above, and then $L_{bol}\sim 4\times
10^{40}$ erg s$^{-1}$. This $L_{bol}$ is a factor of $\sim 2$ lower
than $L_{acc}$, which could be explained by a similarly lower value of the
radiative efficiency (with respect to the standard value of 0.1 used
to estimate $L_{acc}$ in Sect.~\ref{dead}) and/or of  $\dot M_{in}$
(due, e.g., to angular momentum in the flow before being
captured by the MBH, Sect.~\ref{ang}). With a 20\% reflectance, and
the H$\alpha$ upper limit of Sect.~\ref{bc03}, this nucleus would lie
close to the L(2--10 keV)--L$_{H\alpha}$ correlation (Panessa et
al. 2006), at higher X-ray luminosities with respect to the best fit
line but within the observed scatter.

Another observational constraint comes from the fact that thermal
reradiation of an obscured continuum should appear as dust emission in
the IR.  The IRAC data in Fig.~\ref{sed} fit very well to the SED of a
stellar population (Sect.~\ref{bc03}), so that any warm/hot dust must
be negligible. The $8\mu m$ flux (Tab.~\ref{tablesed}) corresponds to
$3.4\times 10^{40}$ erg s$^{-1}$, and any fraction due to dust must be
much less than this. A conservative limit would be that the dust flux
is less than the difference of the fluxes at $8\mu m$ and $5.8\mu m$,
otherwise the SED would depart from the galaxy template. That
difference (0.5 mJy, from Tab.~\ref{tablesed}) corresponds to a
luminosity limit of $<3.1\times 10^{40}$ erg s$^{-1}$, that is a
factor of a few less than $L_{acc}$.  The prevalence of a
non-nuclear origin for the $Spitzer$ IRAC measurements, that refer to
an area of 4 square arcseconds, is indicated also by a color
analysis. $Spitzer$ IRAC colors have been shown to provide a robust
technique for identifying AGNs, i.e., inactive galaxies can be
separated from AGNs in a [3.6]--[4.5] versus [5.8]--[8.0] plot (Stern
et al. 2005). We calculated such colors for the
nucleus of NGC821, using fluxes derived with a local background in
order to better single out the nuclear emission (Tab.~\ref{spizflu}),
and found that it falls well within the region of normal
galaxies. This confirms that stars within the central $1\farcs1$
radius dominate over the IR nuclear emission (Sect.~\ref{bc03}).
Longer wavelength data are crucial to definitely rule out the
possibility of heavily obscured emission. Unfortunately, NGC821 has
not been detected by $IRAS$ (Knapp et al. 1989)\footnote{With IRAS
data just an uncertain $1\sigma$ detection of 0.5 Jy at 100$\mu m$
could be placed [J. Knapp 1994, private communication to NED
(http://nedwww.ipac.caltech.edu/)], corresponding to $\sim 10^{42}$
erg s$^{-1}$.}. $Spitzer$ MIPS imaging has been scheduled and,
possibly together with deeper $HST$/NICMOS F160W images, should be
useful to definitely test the hypothesis of a hidden AGN.

\section{Summary and conclusions}\label{concl}

NGC821 is an extreme example of a quiescent MBH, for which deep
$Chandra$ and VLA observations revealed a very sub-Eddington
nuclear source for the first time. Since NGC821 is relatively nearby,
it is a good testcase to study how accretion proceeds in the very low
luminosity galactic nuclei that are the vast majority in the
local universe (Sect.~\ref{intro}). Our multiwavelength
analysis has shown that:

\begin{enumerate}

\item A radio counterpart to the $Chandra$ nuclear source S2 is
detected at 1.4 GHz, with a flux density of 127 $\mu$Jy. A source is
also marginally (at the $\sim 4 \sigma$ level) detected at 4.8 and 8.4 GHz.
Within current uncertainties, the radio spectral shape of the nuclear
source is relatively flat. Upper limits can be placed at the positions of the
other elongated $Chandra$ sources in the central galactic region (S1 and S4). 

\item The central $\sim 1\farcs1$ radius region is also detected with
$Spitzer$ IRAC, with the emission peak coincident with the position of the
galactic center. Archival $HST$ images taken with WFPC2/F555W, WFPC2/F814W and
NICMOS (H-band) do not show a pointlike source at the galactic center.

\item A dead MBH could be expected from the lack of detection of fuel
at all wavelengths. However, the stellar mass losses in the
circumnuclear region should produce a luminosity of $L_{acc}\sim
$few$\times 10^{41}$ erg s$^{-1}$, if they end in a standard disc
(such a scaled-down AGN would work at $L_{acc}/L_{Edd}\sim $few$\times
10^{-5}$).

\item Disc-like stellar kinematics in the central galactic region,
with the consequent angular momentum of the accretion flow at large
radii, may account for a reduction of the mass accretion rate; MHD
simulations with rotation and cooling by radiation included are needed
to estimate the final luminosity of the flow.

\item Starformation in the accreting material could also reduce the
actual fuel supply to the MBH.  The mid-IR to V-band photometric data
for the central $1\farcs1$ radius region, though, are very well fitted
by the spectral energy distribution of an old and metal rich stellar
population.

\item The upper limit on pointlike 2--10 keV nuclear emission,
together with the 5 GHz ($4.0\sigma$) and 8.4 GHz ($4.2\sigma$)
detections, and the MBH mass known from $HST$ data, are consistent
with the predictions for radiatively inefficient accretion coupled to
a compact, nuclear jet. The relatively flat radio
spectral shape of the nuclear source is also consistent with
being produced by the base of a jet.

\item The extended $Chandra$ sources S1 and S4 
could be due to several knots in a jet, given their
X-ray--to--radio luminosity ratios not too far from those observed for
the knots in the jet of Cen A; more sensitive radio observations are
needed to draw final conclusions on this possibility. If these sources
or the central radio source are jet-like, their mechanical energy
would be a very small fraction (a few$\times 10^{-4}$) of the energy
associated with the rest mass of the material being accreted.

\item The nucleus of NGC821 could be a standard AGN working at
$L_{acc}/L_{Edd}\sim$few$\times 10^{-5}$ but heavily obscured as in a
Compton-thick source, possibly also extended as S2 is. The NICMOS
upper limit on the nuclear emission constrains the reflectance in the
X-rays, but does not rule out this possibility.
Thermal re-radiation by dust is expected in the obscured
scenario: the possible presence of warm/hot dust is already
constrained by the $Spitzer$ mid-IR data, but far-IR observations are
needed to assess the emission from cold dust.

\end{enumerate}

In conclusion, the deep study of the nucleus of NGC821 has shown that
MBHs in the local universe are still emitting, though at an extremely
low level, even in low $L_B$ ellipticals, that are expected to be very
poor of hot gas.  This agrees with the findings of
hydrodynamical simulations that stellar mass losses in the
circumnuclear region can fuel the MBH, even in low $L_B$ galaxies.
The present multiwavelength investigation leaves open the
possibilities that the final stages of accretion are radiatively
inefficient, or that a standard disc in a Compton-thick scaled-down
AGN is present (provided that the radiative efficiency and/or the mass
accretion rate are reduced). Deeper NICMOS and radio observations, together 
with far-IR data, are expected to solve the puzzle of the nature of the
accretion process in this intriguing nucleus.

\acknowledgments

We thank M. Bolzonella for having provided the fitting SEDs for
Sect.~\ref{bc03}. S.P. acknowledges financial support from ASI
(Agenzia Spaziale Italiana) contract I/023/05/0.  Partial support for
this work was provided by the NASA $Chandra$ Guest Observer grant
GO5-6110X and by the $Chandra$ X-ray Center NASA contract NAS8-39073 and
by the Spitzer Cycle-2 program 20371. The data analysis was supported
by the CXC CIAO software and CALDB. We have used NASA NED and ADS
facilities, and have extracted archival data from the Hubble Space
Telescope archive.

\clearpage

\begin{deluxetable}{lccccccccc}
\rotate
\tablecaption{NGC~821: main properties\label{mainlog}}
\tablewidth{0pt}
\tablehead{
\colhead{Type\tablenotemark{a}} &
\colhead{B$^0_T$\tablenotemark{a}} &
\colhead{D\tablenotemark{b}}& 
\colhead{log $L_{B}$} &
\colhead{Size\tablenotemark{a}} &
\colhead{R$_{\rm e}$\tablenotemark{c}} &
\colhead{$\sigma_e$\tablenotemark{d}} &
\colhead{N$_H$\tablenotemark{e}} &
\colhead{$M_{\rm BH}$\tablenotemark{f}} &
\colhead{1$^{\prime\prime}$}     
\\
\colhead{ } &
\colhead{(mag)} &
\colhead{(Mpc)} &
\colhead{($L_{B,\odot}$)} &
\colhead{(arcmin)} &
\colhead{($^{\prime\prime}$,kpc)} &
\colhead{(km~s$^{-1}$)} &
\colhead{(cm$^{-2}$)} &
\colhead{($10^7 M_{\odot}$)}&
\colhead{(pc)}
}
\startdata
 E6 & 11.72 & 24.1 & 10.27 & 2.57x1.62 & 43.9, 5.1 & 209 & 6.2$\times 10^{20}$ & 8.5$\pm 3.5$ & 117 \\
\enddata

\tablenotetext{a}{Type, $B^0_T$ and size from de Vaucouleurs et al. (1991; 
RC3). The size gives the major and minor axis of the D25
ellipse, that is the 25.0 B-mag/square arcsec isophote. The position
angle is $25^{\circ}$ (RC3).}

\tablenotetext{b}{Distance D from Tonry et al. (2001).}

\tablenotetext{c}{Effective radius R$_{\rm e}$ in the R-band (from 
Soria et al. 2006b).}

\tablenotetext{d}{Effective stellar velocity dispersion (averaged 
over $R_e$) from Pinkney et al. (2003).}

\tablenotetext{e}{Galactic hydrogen column density
(Dickey \& Lockman 1990).}

\tablenotetext{f}{Gebhardt et al. (2003) report a value of
3.7$^{+2.4}_{-0.8}\times 10^7\,M_{\odot}$, later revised to the value
given here (Richstone et al., astro-ph/0403257) that is considered
more reliable (Gebhardt, K. 2006, private communication).}

\end{deluxetable}

\clearpage

\begin{table}[ht]
\caption{The $Chandra$ source S2}\label{tabspec}
\begin{center}
\begin{tabular}{lccc}
\hline\hline
Position (J2000)                 & RA          &    Dec   & Ref.     \\
$HST$ (galactic center) & \hskip-1.truecm $02^h$ $08^m$ $21^s$\hskip-0.1truecm.13 & $+10^{\circ}$ $59^\prime$ $41\farcs8$ &P07 \\
$Chandra$               &  \hskip-1.truecm $02^h$ $08^m$ $21^s$\hskip-0.1truecm.10 & $+10^{\circ}$ $59^\prime$ $41\farcs6$& P07 \\
VLA                     & \hskip-1.truecm $02^h$ $08^m$ $21\rlap{.}''174$         & $+10^\circ$   $ 59' $     $41\rlap{.}''4$& Sect.~\ref{vla}\\
$Spitzer$            &  \hskip-1.truecm $02^h$ $08^m$ $21^s$\hskip-0.1truecm.11 &$+10^{\circ}$ $59^\prime$ $42\farcs0$ & Sect.~\ref{spiz} \\
\hline
Size\tablenotemark{a} & $1\farcs7\times 1\farcs5$ & \\
\hline
Spectral analysis of $Chandra$ data: & & \\
Net counts\tablenotemark{b} &  246$\pm 16$   &     \\
Model $wabs(pow)$ : & & \\
N$_H$ (10$^{21}$ cm$^{-2}$)&  $<0.5$& \\
$\Gamma$ &  $1.49_{-0.13}^{+0.14}$& \\
$\chi^2/$dof & 11.5/9 & \\
L(0.3--8 keV)/$10^{38}$ erg s$^{-1}$ &  6.0 & \\
L(2--10 keV)/$10^{38}$ erg s$^{-1}$ &  3.8 & \\
Model  $wabs(pow+gauss)$\tablenotemark{c} : &  & \\
N$_H$ (10$^{21}$ cm$^{-2}$)&  $<0.5$ & \\
$\Gamma$ &  $1.55_{-0.17}^{+0.14}$& \\
EW   (keV)    & 0.9 ($<2.0$) & \\ 
$\chi^2/$dof & 11.2/8 & \\
L(0.3--8 keV)/$10^{38}$ erg s$^{-1}$ & 6.3 & \\
L(2--10 keV)/$10^{38}$ erg s$^{-1}$ &  5.0 & \\
\tbsp
Pointlike emission\tablenotemark{d} $\,$ within S2: & & \\
L(0.3--8 keV)/$10^{38}$ erg s$^{-1}$ & $<2.8$ & \\
L(2--10 keV)/$10^{38}$ erg s$^{-1}$ &  $<1.8$ & \\
\hline\hline
\end{tabular}

\end{center}
\tablenotetext{a}{\phantom{}Length of semi-major and semi-minor axes of the ellipse 
describing the source shape, derived by the CIAO task $wavdetect$ (P07).}

\tablenotetext{b}{From a total net exposure time of 226 ksec.}

\tablenotetext{c}{The energy of the gaussian emission line has been fixed at 6.4 keV.}

\tablenotetext{d}{Calculated as described in P07.}

\tablecomments{The column density $N_H$ is in addition to the Galactic
one. Errors give the 68\% confidence
interval for one interesting parameter.}
\end{table}

\clearpage

\begin{table}[ht] 
\caption{The $Chandra$ sources S1 and S4}\label{s1-s4}
\begin{center}
\begin{tabular}{lcc} 
\hline\hline
           &   S1   &   S4  \\
\hline 
Size (ellipse)\tablenotemark{a} & $2\farcs5\times 1\farcs5$ & \hskip0.3truecm
$1\farcs2\times 1\farcs0$\\
Net counts\tablenotemark{b} & 178$\pm 14$ & \hskip0.3truecm 82$\pm 9$   \\
Model $wabs(pow)$ : & & \\
N$_H$ (10$^{21}$ cm$^{-2}$)& $1.7_{-1.0}^{+0.8}$& \hskip0.3truecm $3.0_{-1.5}^{+1.3}$\\
$\Gamma$ & $1.80_{-0.23}^{+0.35}$ & \hskip0.3truecm $2.28_{-0.34}^{+0.52}$ \\
$\chi^2/\nu $ & 7.3/6 &\hskip0.3truecm 2.1/5 \\
L(0.3--8 keV)/$10^{38}$ erg s$^{-1}$ & 4.7 &\hskip0.3truecm 1.9 \\ 
\hline\hline
\end{tabular} 

\end{center} 
\tablenotetext{a}{\phantom{}Length of semi-major and semi-minor axes of the ellipse 
describing the source shape, derived by the CIAO task $wavdetect$ (P07).}

\tablenotetext{b}{From a total net exposure time of 226 ksec.}

\tablecomments{The column density $N_H$ is in addition to the Galactic one (see P07). Errors give the 68\% confidence
interval for one interesting parameter.}
\end{table}

\clearpage

\begin{table}
\caption{$Spitzer$ fluxes from an area of 4 square arcseconds centered on the
nucleus of NGC~821.\label{spizflu}}
\begin{center}
\begin{tabular}{ccc}
\hline
Band   &     Flux density (mJy) &       Flux density (mJy)\\
       &    local bkgd          &      field bkgd        \\
\hline\hline
3.6 $\mu m$ &     2.4 $\pm$ 0.5 &      5.6 $\pm$ 0.5\\
4.5  $\mu m$ &    1.1 $\pm$ 0.4 &      3.1 $\pm$ 0.4\\
5.8  $\mu m$ &    0.7 $\pm$ 0.2 &      1.8 $\pm$ 0.2\\
8   $\mu m$ &     0.4 $\pm$ 0.3 &      1.3 $\pm$ 0.2\\    
\hline\hline
\end{tabular}
\end{center}
\tablecomments{Fluxes are given for two choices of the background:
estimated from a surrounding annulus (local)
and from an off-source circle (field), see Sect.~\ref{spiz}. }
\end{table}

\clearpage

\begin{deluxetable}{ccrcc}
\rotate
\tablecaption{Spectral energy distribution of the nucleus of NGC~821.\label{tablesed}}
\tabletypesize{\scriptsize}
\tablewidth{0pt}
\tablehead{
\colhead{Band} &
\colhead{Flux Density} & 
\colhead{$\nu L_{\nu}$ (erg s$^{-1})$} & 
\colhead{Region ($^{\prime\prime}$)} &
\colhead{Instrument}
}
\startdata
1.4 GHz     &  $127\,\,\mu$Jy & $1.2\times 10^{35}$   & $1.3\times 1.3$\tablenotemark{a}   & VLA 
(Sect.~\ref{vla}) \\
4.8 GHz     &  ($80\,\,\mu$Jy)\tablenotemark{b} & ($2.7\times 10^{35}$) &$\,\,\,\,\, 4.1 \times 4.0$\tablenotemark{a}  & $^{\prime\prime}$ \\
8.4 GHz     &  ($71\,\,\mu$Jy)\tablenotemark{b} & ($4.1\times 10^{35}$) & $\,\,\,\,\, 2.7 \times 2.2$\tablenotemark{a}   & $^{\prime\prime}$ \\
15 GHz      &  $<1.5$ mJy   & $<1.5\times 10^{37}$   & R=0.15 circle & VLA ($5\sigma$ limit; Nagar et al. 2005) \\
8   $\mu m$ &  1.3 $\pm$ 0.2 mJy & $3.4\times 10^{40}$  & R=$1.1$ circle & $Spitzer$ IRAC (Sect.~\ref{spiz}) \\
5.8 $\mu m$ &  1.8 $\pm$ 0.2 mJy & $6.5\times 10^{40}$  & $^{\prime\prime}$ & $^{\prime\prime}$ \\
4.5 $\mu m$ &  3.1 $\pm$ 0.4 mJy & $1.4\times 10^{41}$  & $^{\prime\prime}$ & $^{\prime\prime}$ \\
3.6 $\mu m$ &  5.6 $\pm$ 0.5 mJy & $3.2\times 10^{40}$  &  $^{\prime\prime}$ & $^{\prime\prime}$ \\
1.6 $\mu m$ &  $<0.01$ mJy\tablenotemark{c} & $<1.4\times 10^{39}$   & R=$0.02$ circle & NICMOS (Ravindranath et al. 2001) \\
I ($0.90\mu m$) & 7.45$^{+0.72}_{-0.66}$ mJy & $1.9\times 10^{42}$  & R=$1.1$ circle & WFPC2 F814W (Sect.~\ref{opt})\\
 $^{\prime\prime}$ & $<0.19$ mJy\tablenotemark{c} & $<4.3\times 10^{40}$ &  R=$0.046$ circle & $^{\prime\prime}$ \\
R ($0.70\mu m$) & 5.47$^{+0.53}_{-0.48}$ mJy & $1.7\times 10^{42}$ & R=$1.1$ circle & INT (Sect.~\ref{opt})\\
V ($0.55\mu m$) & 3.54$^{+0.34}_{-0.31}$ mJy & $1.3\times 10^{42}$ & $^{\prime\prime}$ & WFPC2 F555W (Sect.~\ref{opt})\\
 $^{\prime\prime}$ & $<0.075$ mJy\tablenotemark{c} & $<2.8\times 10^{40}$ & R=$0.046$ circle & $^{\prime\prime}$ \\
$2.4\times 10^{17}$ Hz       & $\,\,\,\,\,<6.5\times10^{-7}$ph.cm$^{-2}$s$^{-1}$keV$^{-1}$  & $\,\,\, <7.2\times 10^{37}$ & $1\times 1$  & $Chandra$ ACIS-S\tablenotemark{d} $\,$ (Sect.~\ref{endistr}) \\
$4.8\times 10^{17}$ Hz       & $\,\,\,\,\,<2.0\times 10^{-7}$ph.cm$^{-2}$s$^{-1}$keV$^{-1}$  & $\,\,\, <8.7\times 10^{37}$  & $^{\prime\prime}$  & $^{\prime\prime}$\\
$2.4\times 10^{18}$ Hz       & $\,\,\,\,\,<1.3\times 10^{-8}$ph.cm$^{-2}$s$^{-1}$keV$^{-1}$  & $\,\,\, <1.7\times 10^{38}$ & $^{\prime\prime}$ & $^{\prime\prime}$\\
\enddata

\tablenotetext{a}{"Region" gives here the beam size; the
position angle of the beam is -56$^{\circ}$ at 4.8 and 8.4 GHz.}

\tablenotetext{b}{The 4.8 GHz is a $4.0\sigma$ detection, and the 8.4 GHz is
a $4.2\sigma$ detection.}

\tablenotetext{c}{Upper limits to any "AGN" flux density (the brightness profile is
consistent with that of the normal galactic stellar emission).}

\tablenotetext{d}{$3\sigma$ upper limits on pointlike nuclear emission (P07).}

\end{deluxetable}

\clearpage

\clearpage

\clearpage

\clearpage

\end{document}